\documentclass[pra,aps,twocolumn,10pt,superscriptaddress]{revtex4-2}%

\usepackage{mathrsfs}
\usepackage[scr=rsfs,cal=boondox]{mathalfa}
\usepackage[normalem]{ulem}
\usepackage{amsmath}
\usepackage{amsfonts}
\usepackage{amssymb}
\usepackage{graphicx}%
\usepackage[usenames]{color}
\usepackage{color}
\usepackage{xcolor}
\usepackage{appendix}
\DeclareMathAlphabet{\mathcal}{OMS}{cmsy}{m}{n}
\newcommand{\pd}[2]{\frac{\partial #1}{\partial #2}}
\newcommand{\der}[2]{\frac{d #1}{d #2}}
\newcommand{\pdd}[2]{\frac{\partial^2 #1}{\partial #2^2}}

\definecolor{amber}{rgb}{1.0, 0.49, 0.0}

\usepackage{hyperref}
\hypersetup{
    colorlinks,
    linkcolor={red!70!black},
    citecolor={green!65!black},
    urlcolor={blue!80!black}
}

\providecommand{\U}[1]{\protect\rule{.1in}{.1in}}
\begin{document}
\title{%A one dimensional model for ultracold collisions of a neutral atom with a trapped ion
Ultracold collisions of a neutral atom with a trapped ion in 1D
}
\author{Seth T. Rittenhouse}
\affiliation{Department of Physics, the United States Naval Academy, Annapolis, MD, 21402, USA}
\affiliation{ITAMP, Center for Astrophysics $|$ Harvard $\&$ Smithsonian, Cambridge, USA}
\affiliation{Institute for Theoretical Physics, Institute of Physics, University of Amsterdam, Science Park 904, 1098 XH Amsterdam, the Netherlands}
\author{Lorenzo Oghittu}
\affiliation{Institute for Theoretical Physics, Institute of Physics, University of Amsterdam, Science Park 904, 1098 XH Amsterdam, the Netherlands}
\affiliation{QuSoft, Science Park 123, 1098 XG Amsterdam, the Netherlands}
\author{Arghavan Safavi-Naini}
\affiliation{Institute for Theoretical Physics, Institute of Physics, University of Amsterdam, Science Park 904, 1098 XH Amsterdam, the Netherlands}
\affiliation{QuSoft, Science Park 123, 1098 XG Amsterdam, the Netherlands}
\author{Rene Gerritsma}
%\affiliation{Institute for Theoretical Physics, Institute of Physics, University of Amsterdam, Science Park 904, 1098 XH Amsterdam, the Netherlands}
\affiliation{Van der Waals-Zeeman Institute, Institute of Physics, University of Amsterdam, 1098 XH Amsterdam, the Netherlands}
%\author{Angela D. Graf}
%\affiliation{Department of Physics and Astronomy, Trinity University, San Antonio Texas 78212, USA}
\author{Nirav P. Mehta}
\affiliation{Department of Physics and Astronomy, Trinity University, San Antonio Texas 78212, USA}
\affiliation{Institute for Theoretical Physics, Institute of Physics, University of Amsterdam, Science Park 904, 1098 XH Amsterdam, the Netherlands}

\begin{abstract}
We present a fully quantum mechanical description of a free $^6$Li atom scattering from a trapped $^{171}$Yb$^+$ ion in one dimension.  By reformulating the system in polar coordinates and employing the adiabatic representation, we extract a set of coupled adiabatic potentials representing the atom interacting with the ion in different trap states. In an approach similar to quantum defect theory (QDT), we leverage the vast difference in energy scale between the interaction, the trap, and the scattering energy to encapsulate the short-range atom-ion scattering behavior in a single phase parameter.  The presence of trapped $({}^{171}\text{Yb}^6\text{Li})^+$ molecular-ion states leads to a series of roughly evenly spaced resonances in the scattering cross section. The predicted distribution of resonances at low collision energies is at odds with the expectation of quantum chaos and the Bohigas-Giannoni-Schmit (BGS) conjecture. 

\end{abstract}
\maketitle

\section{Introduction}
Hybrid systems consisting of a trapped ion embedded in an ultracold gas of neutral atoms offer exciting prospects in quantum science~\cite{zipkes2010trapped,schmid2010dynamics,meir2016dynamics,tomza2019cold,Cote2016chapter,lous2022}. Such systems benefit from the ultracold temperatures achieved in atomic gases and the ability to individually address the spatially localized trapped ions. Atom-ion interactions serve to entangle the constituent systems, paving the way for several exciting experimental advances, including the observation of atom-ion collisions~\cite{grier2009observation,harter2012single}, magnetic Feshbach resonances~\cite{Weckesser2021oof}, realizations of sympathetic~\cite{zipkes2010trapped,haze2018cooling,Feldker2020bgc} and swap~\cite{mahdian2021direct} cooling of the ion by the cold buffer gas, and the study of ultracold atom-ion chemistry~\cite{ratschbacher2012chemical,sikorsky2018quantum,Katz2022qld}. 

A comprehensive understanding of atom-ion hybrid systems hinges on the development of a robust collision theory that is capable of incorporating the confinement of the ion by a radio-frequency (RF) Paul trap. To date, progress has been made considering classical trajectories~\cite{Cetina2012flt}, which revealed that micromotion heating of the ion can be minimized when the atomic mass $m_\mathrm{a}$ is chosen to be much smaller than the mass of the ion, $m_\mathrm{i}$. More recently, classical calculations have helped interpret experiments observing trap-assisted formation of long-lived atom-ion complexes~\cite{pinkas2023trap,Hirzler2023}. Moreover, classical trajectories exhibit signatures of chaos~\cite{pinkas2024chaotic,liang2025trap} linked to such states reminiscent of the “sticky collisions” predicted~\cite{mayle2012statistical,mayle2013scattering,Croft:2017pra} and observed~\cite{gregory2019sticky} %\NMcomment{(other observations?)} 
in ultracold molecular gases.

While rigorous quantum-mechanical theories considering collisions of atoms and ions in free space have been developed, both at higher thermal energies~\cite{delos1981theory} and in the ultracold regime~\cite{idziaszek2009quantum,idziaszek2011multichannel}, there is yet to appear a fully quantum model that incorporates the ion confinement. Even at the level of the so-called secular approximation in which ion micromotion is neglected, the harmonic trapping potential significantly complicates the collision physics: the trap center acts as an infinitely massive third particle, and one must consider the system as an effective three-body problem.

As a consequence, new dynamical processes---akin to confinement-induced resonances~\cite{melezhik2016confinement}---arise that are only present when the ion confinement is considered. The most significant of these, and the focus of this work, is the formation of a trapped atom-ion complex during the collision. The binding energy from the formation of the molecular ion propels its center of mass into a high-lying state of the oscillator, and the process is resonantly enhanced if such a state lies near the collision energy. Of particular interest in this work is the density and distribution of such confinement resonances and their possible relation to quantum chaos. 

Because classical calculations of atom trajectories in comparable 1D systems have revealed signatures of chaos~\cite{pinkas2024chaotic,liang2025trap}, the 
statistical properties of the corresponding quantum mechanical resonances are of keen interest. According to a conjecture by Bohigas, Giannoni, and Schmit~\cite{bohigas1984characterization}, distributions of the nearest-neighbor energy level spacings in the limit of strong chaos should obey the Wigner-Dyson (WD) distribution~\cite{wigner:characteristic_1955,wigner:characteristics_1957,Brody:1981rmp,Brody:1973LettNuovoCimento}. We investigate the structure of resonances that arise from atom-ion complex states from a fully quantum-mechanical calculation.

We propose a simple, %yet fully quantum mechanical
one-dimensional model for the collision of a free $^6\text{Li}$ atom with a trapped $^{171}\text{Yb}^+$ ion. We neglect trap micromotion, and assume harmonic confinement for the ion. We solve the Schr\"odinger equation for the reactance $K$-matrix, and compute both elastic and inelastic cross sections for the atom-ion collision, which can take the ion from one oscillator state to another. 
%\NMcomment{We still need a "paper outline", but we can add that after we have the structure ironed out.}

This manuscript is arranged as follows. In Section~\ref{sec:methods} we detail the methods used in the work, including the adiabatic hyperspherical representation, the parameterization of the short-range 2-body interaction in terms of a short-range phase, and the implementation of long-range scattering boundary conditions.  In Section~\ref{sec:results} we present and analyze the results including the adiabatic potentials, and the elastic and inelastic scattering cross sections.  In Section~\ref{sec:summary} we summarize the results and discuss open questions and future endeavors. 
\section{Methods}
\label{sec:methods}
We consider an ion in a linear Paul trap with secular frequency $\omega$, interacting with a ground-state atom that does not feel the Paul trap potential. We assume that both the atom and the ion are confined to one spatial dimension. Although a 1D model is unlikely to be fully quantitatively accurate, similar models have been used by several authors in classical simulations to great effect~\cite{Cetina2012flt,pinkas2024chaotic,trimby2022buffer}. The Hamiltonian for this system is given by
\begin{equation}
\hat{H}=\hat{T}_\mathrm{i}+\hat{T}_\mathrm{a}+V_\mathrm{trap}\left(  x_\mathrm{i}\right)  +V_\mathrm{int}\left(
x_\mathrm{a}-x_\mathrm{i}\right)  , \label{Eq:H_ai}%
\end{equation}
where $x_{\mathrm{i\left(  a\right)  }}$ and $\hat{T}_{\mathrm{i\left(  a\right)  }}$ are
respectively the position and kinetic energy operator of the ion (atom). 
%We choose to work in oscillator units corresponding to the secular motion of the ion, where $V_{trap}\left(  x_{i}\right)  = x_{i}^{2}/2$.  The atom and ion coordinates are scaled by the oscillator length $\sqrt{\hbar/m_i \omega}$ where $m_i$ is the ion mass and $\omega$ is the secular frequency of the trap.
\ Here, $V_\mathrm{trap}\left(  x_\mathrm{i}\right)  =m_\mathrm{i}\omega^{2}x_\mathrm{i}^{2}/2$ is the
trapping potential of the linear Paul trap acting on the ion.
\ For the
purposes of this work, we neglect trap micromotion and focus purely on
the harmonic secular portion of the potential~\cite{leibfried2003quantum}.
\ The atom-ion interaction potential, $V_\mathrm{int}\left( r\right)$, is that for
an $s$-wave atom interacting with an ion which we model as~\cite{tomza2019cold}
\begin{equation}
V_\mathrm{int}\left(  r\right)  =-\frac{C_{4}}{r^{4}}+\frac{C_{6}}{r^{6}},
\label{Eq:V_int}%
\end{equation}
where $r=\left\vert x_\mathrm{a}-x_\mathrm{i}\right\vert $ is the atom-ion separation
distance. The attractive $C_4$ term represents the interaction between the ion and the induced dipole, while the $C_6$ term models the short-range interaction and core repulsion. 
%The\ $C_{4}$ and $C_{6}$ terms\ can be simply thought of as the
%interaction between the ion and the induced dipole and quadrupole of the atom, respectively. 
For the $^{171}\text{Yb}^+ + ^{6}\text{Li}$ system, we choose parameters in atomic units (a.u.) that have been previously used in 1D classical calculations: $C_{4}=82$~a.u.
and $C_{6}=2989.4$~a.u~\cite{trimby2022buffer}. Note that this choice of $C_6$ is somewhat arbitrary and unphysical.  It is chosen here to fix the low-energy classical turning point to be in agreement with the classical simulations of ~\cite{trimby2022buffer}.  

To treat the kinetic energy of both the atom and ion on an equal footing, we
transform into a mass-scaled set of coordinates%
\begin{align}
x    =\sqrt{\frac{m_\mathrm{i}}{\mu}}x_\mathrm{i},\label{Eq:jac_coords}\quad
y    =\sqrt{\frac{m_\mathrm{a}}{\mu}}x_\mathrm{a},
\end{align}
where $m_{\mathrm{i\left(  a\right) } }$ is the mass of the ion (atom) and the
effective reduced mass $\mu=\sqrt{m_\mathrm{i}m_\mathrm{a}}$ is chosen to preserve the
volume element in the 2D space, i.e. $dx_\mathrm{i}dx_\mathrm{a}=dxdy$. In these
mass-scaled coordinates the total kinetic energy is simply%
%\begin{align*}
%\hat{T}_\mathrm{i}+\hat{T}_\mathrm{a} &=-\frac{\hbar^{2}}%{2\mu}\nabla^{2},\\
%\nabla^{2}  &  =\frac{\partial^{2}}{\partial %x^{2}}+\frac{\partial^{2}%
%}{\partial y^{2}}.
%\end{align*}
\begin{equation}
\hat{T}_\mathrm{i}+\hat{T}_\mathrm{a} =-\frac{\hbar^{2}}{2\mu}\nabla^{2};\;\;\;\;\;
\nabla^{2}    =\frac{\partial^{2}}{\partial x^{2}}+\frac{\partial^{2}%
}{\partial y^{2}}.
\end{equation}
For this work, we will make one more transformation and rewrite the system in
polar coordinates:%
\begin{align*}
x  & =R\cos\theta,\quad\\
y  & =R\sin\theta.
\end{align*}
Here, the “hyperradius” $R=\sqrt{x^{2}+y^{2}}$ can be thought of as the RMS size of the atom-ion
system as measured from the center of the ion trap. In contrast to the normal
interpretation of polar coordinates, the “hyperangle” $\theta$ does not correspond to a
standard spatial angle, but can be thought of as parameterizing the
configuration of the atom and ion. For example, if $\theta=\pi/2$, the ion is at
the center of the trap and the atom is at the maximum allowed distance from
the trap center, and if $\theta=0$, the reverse is true. At the specific
intermediate angle $\theta=\theta_{c}=\tan^{-1}\left(  \sqrt{m_\mathrm{a}%
/m_\mathrm{i}}\right)$, the two are on top of one another. The full Hamiltonian in
polar coordinates is given by%
\begin{align}
\hat{H}=&-\frac{\hbar^{2}}{2\mu}\frac{1}{R}\pd{}{R}R\pd{}{R}-\frac{\hbar^{2}}{2\mu R^{2}}\frac{\partial^{2}}{\partial\theta^{2}}%
+\frac{1}{2}\mu\omega^{2}R^{2}\cos^{2}\theta\label{Eq:H_polar}\nonumber\\
&  +V_\mathrm{int}\left(  \sqrt{\beta}R\sin\left(  \theta-\theta_{c}\right)  \right)
,
\end{align}
where $\beta=\left(  m_\mathrm{i}+m_\mathrm{a}\right)  /\sqrt{m_\mathrm{i}m_\mathrm{a}}.$

\subsection{The adiabatic representation}
\label{sec:adiabaticrep}
Potential energy curves are extracted by treating the radius $R$ as an adiabatic parameter~\cite{macek1968pas}. To do this, we consider the adiabatic Hamiltonian $\hat
{H}_\mathrm{ad}$ defined as the part of $\hat{H}$ that remains if $R$ is held
constant:%
\begin{align}
\hat{H}_\mathrm{ad}   =&-\frac{\hbar^{2}}{2\mu R^{2}}\frac{\partial^{2}}%
{\partial\theta^{2}}+\frac{1}{2}\mu\omega^{2}R^{2}\cos^{2}\theta
\label{Eq:H_ad} \nonumber\\
&  +V_\mathrm{int}\left(  \sqrt{\beta}R\sin\left(  \theta-\theta_{c}\right)  \right)
.
\end{align}
\ The solution to the Schr\"{o}dinger equation (SE) $\hat{H}\Psi=E\Psi$ is
then expanded in eigenfunctions of the adiabatic Hamiltonian, i.e.%
\begin{align}
\sqrt{R}\Psi\left( R,\theta\right)   &  =\sum_{n}F_{n} \left(R\right)
\Phi_{n}\left(  R;\theta\right)  ,\label{Eq:ad_exp}\\
\hat{H}_\mathrm{ad}\Phi_{n}\left( R;\theta\right)   &  =U_{n}\left( R\right)
\Phi_{n}\left( R;\theta\right)  . \label{Eq:ad_SE}%
\end{align}
Inserting Eq.~\eqref{Eq:ad_exp} into the SE creates a set of coupled 1D
Schr\"{o}dinger equations for the radial function $F_{n}\left(  R\right)  $:%
\begin{align}
EF_{n}\left(  R\right)    & =\frac{-\hbar^{2}}{2\mu}\left(  \frac{d^{2}%
}{dR^{2}}+\frac{1}{4R^{2}}\right)  F_{n}\left(  R\right)  +U_{n}\left(
R\right)  F_{n}\left(  R\right) \nonumber \\
 -&\frac{\hbar^{2}}{2\mu}\sum_{m}\left(  Q_{nm}\left(  R\right)
+2P_{nm}\left(  R\right)  \frac{d}{dR}\right)  F_{m}\left(  R\right)\label{Eq:rad_SE}
,
\end{align}
Note that the extra $1/(4R^{2})$ term is due to the extra factor of $\sqrt{R}$
included on the left-hand-side of Eq.~(\ref{Eq:ad_exp}), which also serves to
remove the first derivative terms from the SE. Here, the matrix elements of
the non-adiabatic $P$- and $Q$-matrices are defined as%
\begin{align}
P_{nm}\left(  R\right)   &  = \left\langle \Phi_{n}\left(  R;\theta\right)
\middle\vert \pd{\Phi_{m}\left(
R;\theta\right) }{R} \right\rangle ,\label{eq:pmat}\\
Q_{nm}\left(  R\right)   &  = \left\langle \Phi_{n}\left(  R;\theta\right)
\middle\vert \pdd{\Phi_{m}\left(
R;\theta\right) }{R} \right\rangle, \label{qmat}
\end{align}
where the integrals are taken over the angular degree of freedom. Note that the $Q$-matrix can be computed in terms of the $P$-matrix elements through the matrix relation
\begin{equation}
    %Q_{nm}=[P^{2}]_{nm}+\frac{\partial}{\partial R}P_{nm}.\label{Eq:QviaP}
    \mathbf{Q}(R)=\mathbf{P}^{2}(R)+\der{\mathbf{P}(R)}{R}\label{Eq:QviaP}
\end{equation}
It is often inadvisable to use this form, as a large number of adiabatic channels
may be required to accurately describe $\mathbf{P}^{2}$. However, in this case
the $P$-matrix elements fall off away from the diagonal sufficiently quickly for
fast convergence of the square with respect to the number of included channels.

Physically, we can consider the different potentials, $U_{n}(R)$, as scattering channels that are coupled by the non-adiabatic $P$- and
$Q$-matrices.  Off-diagonal elements of $P$ and $Q$ lead to inelastic processes, and the diagonal elements $Q_{nn}(R)$ serve as a correction to the adiabatic potential energy functions, giving a set of effective potentials defined as
\begin{equation}
U_{\mathrm{eff},n}(R)=U_n(R)-\frac{\hbar^2}{2\mu}\left( Q_{nn}(R)+\frac{1}{4R^2}\right).
\end{equation}
The effective potentials are meaningfully interpreted as one-dimensional potential energy functions in each channel. Moreover, including the diagonal correction at long range is essential for obtaining the correct threshold scattering behavior in each collision channel. It is straightforward to show that $\mathbf{P}$ is antisymmetric by demanding the adiabatic eigenfunctions $\Phi_n(R;\theta)$ remain normalized at all $R$. When $R\rightarrow \infty$, we expect both $P_{mn}(R)$ and $Q_{mn}(R)$ to vanish, and some insight is gained by examining the adiabatic potentials $U_n(R)$ by themselves in this limit. 
When $R\gg a_\mathrm{ho}$ (where $a_\mathrm{ho}$ is the ion oscillator length), we expect the ion to be in a trap state with the atom free. Thus, we expect the potentials will approach finite thresholds at the ion oscillator energies. Each potential can then be interpreted as that of a free atom interacting with an ion in the corresponding oscillator state. 
Solving Eq.~(\ref{Eq:rad_SE}) for elastic and
inelastic multi-channel scattering amplitudes gives full access to the
complete dynamics of the system.
%Note in Eq.~(\ref{Eq:rad_SE}) that the $P$-matrix acts only with the derivative of the radial wavefunction.  In contrast, the Q-matrix is more akin to a standard potential matrix, whose off diagonal elements directly couple different channels.  The diagonal of the Q-matrix contributes directly as an additional potential and is often referred to as the diagonal correction.  Combined with the adiabatic potentials $U_n(R)$, this creates a set of effective potentials 
%$$U_{eff,n}(R)=U_n(R)-\frac{\hbar^2}{2\mu}Q_{nn}(R)$$ 
%which are often the physically meaningful potentials.

We calculate the $P$-matrix via the Feynman-Hellmann theorem (FHT) in which
matrix elements of the derivatives of the channel function are replaced by
matrix elements of derivatives of the adiabatic Hamiltonian,%
\[
P_{nm}\left(  R\right)  =\frac{\left\langle \Phi_{n}\left(  R;\theta\right)
\left\vert \frac{\partial\hat{h}_\mathrm{ad}}{\partial R}\right\vert \Phi_{m}\left(
R;\theta\right)  \right\rangle }{\varepsilon_{m}\left(  R\right)
-\varepsilon_{n}\left(  R\right)  }%
\]
where $\hat{h}_\mathrm{ad}=R^{2}\hat{H}_\mathrm{ad}$ and $\varepsilon_{n}\left(  R\right)
=R^{2}U_{n}\left(  R\right)  $. Note that here we have chosen to multiply the
adiabatic Hamiltonian by $R^{2}$ whereby the radial derivative in the FHT eliminates the angular kinetic energy terms. Inserting the adiabatic
Hamiltonian from Eq. (\ref{Eq:H_ad}) yields%
\begin{align}
P_{nm}\left(  R\right)     =\frac{1}{R}\Bigg[  4\frac{\left\langle \Phi
_{n}\left(  R;\theta\right)  \left\vert \frac{1}{2}\mu\omega^{2}R^{2}\cos
^{2}\theta\right\vert \Phi_{m}\left(  R;\theta\right)  \right\rangle }%
{U_{m}\left(  R\right)  -U_{n}\left(  R\right)  }& \label{Eq:PElem} \nonumber\\
-  2\frac{\left\langle \Phi_{n}\left(  R;\theta\right)  \left\vert
V_\mathrm{int}\left(  \sqrt{\beta}R\sin\left(  \theta-\theta_{c}\right)  \right)
\right\vert \Phi_{m}\left(  R;\theta\right)  \right\rangle }{U_{m}\left(
R\right)  -U_{n}\left(  R\right)  } & \Bigg] 
\end{align}
and the matrix $\mathbf{Q}$ is then obtained using Eq.~(\ref{Eq:QviaP}).

In practice, Eq.~(\ref{Eq:ad_SE}) is solved numerically by expressing the adiabatic channel functions as a linear combination of b-splines~\cite{deboor}.  The basis must be flexible enough to represent oscillations in the harmonic potential centered at $\theta=\pm\pi/2$ as well as in the deep atom-ion interaction centered at $\theta = \theta_c$. The simplest basis that can do this at small to moderate values of $R$ is obtained by a linear knot sequence. Numerical calculations of $U_n(R)$, $P_{mn}(R)$ and $Q_{mn}(R)$ are matched smoothly onto analytical expressions for their asymptotic forms, derived in App.~\ref{appen:asymp}. Complications specific to the atom-ion problem are discussed in Sec.~\ref{sec:results}. The set of coupled radial equations shown in Eq.~\ref{Eq:rad_SE} is solved using finite element $\mathbf{R}$-matrix propagation~\cite{burke1999theoretical}.

\subsection{Short-Range Boundary Conditions}
\label{sec:SRBCs}
Solving Eq.~\eqref{Eq:ad_SE} still poses a considerable challenge. One of
the main impediments is the depth of the atom-ion potential. For instance,
in the case of scattering between $\text{Yb}^{+}$ and $\text{Li}$ with a secular frequency on the order of
$\omega\sim2\pi\times1$ MHz, the depth of the potential is of order $10^{6}\hbar\omega$~\cite{tomza2019cold}. Meanwhile, the collision energy is an order of magnitude smaller than an oscillator unit. It is nearly impossible in any
standard numerical approach to describe the wavefunction over such vast energy scales to obtain accurate low energy scattering amplitudes.
 To avoid this difficulty, we develop an approach that leverages these
energy scale disparities as a feature by noting that at short range, the
potential depth is much greater than any feasible scattering energy scale for
an ultra-cold system. As such, we can treat the phase accumulated in the deep short-ranged potential to be effectively independent of energy, taking its zero-energy value as a constant. The challenge,
then, is to connect the short-range, zero-energy solutions to the solution
spanning the full range of angles, $0\leq\theta<2\pi.$

We overcome that challenge here by noting that in the limit of $R\left\vert
\theta-\theta_{c}\right\vert \ll R^{\ast}=\sqrt{2\mu_\mathrm{2b}C_{4}/\hbar^{2}}%
$ (deep in the interaction potential) the adiabatic Hamiltonian reduces to
exactly that of the 2-body atom-ion system in free-space:%
\begin{align}
&  \hat{H}_\mathrm{ad}\underset{\theta\longrightarrow\theta_{c}}{\longrightarrow}%
\hat{H}_\mathrm{2b},\label{Eq:H_ad_limit} \nonumber\\
H_\mathrm{2b}  &  =\frac{-\hbar^{2}}{2\mu_\mathrm{2b}}\frac{d^{2}}{dr^{2}}+V_\mathrm{int}\left(
r\right)  ,
\end{align}
where $r\approx\sqrt{\beta}R\left\vert \theta-\theta_{c}\right\vert $ and
$\mu_\mathrm{2b}=m_\mathrm{i}m_\mathrm{a}/\left(  m_\mathrm{i}+m_\mathrm{a}\right)  $ is the 2-body reduced mass.
Thus, in the strongly interacting limit the angular channel functions
$\Phi_{n}\left(  R;\theta\right)$ approximately solves the zero-energy
atom-ion Schr\"{o}dinger equation%
\begin{equation}
\hat{H}_\mathrm{ad}\Phi_{n}\left(  R;\theta\right)  =U_{n}\Phi_{n}\left(
R;\theta\right)  \underset{r\longrightarrow0}{\longrightarrow}H_\mathrm{2b}%
\psi\left(  r\right)  =0, \label{Eq:2body_ZESE}%
\end{equation}
where $\psi\left(  r\right)  $ is the solution to the 2-body ion-atom
Schr\"{o}dinger equation in the absence of the ionic trapping potential.

\begin{figure}[ptb]
\includegraphics[width=\linewidth]{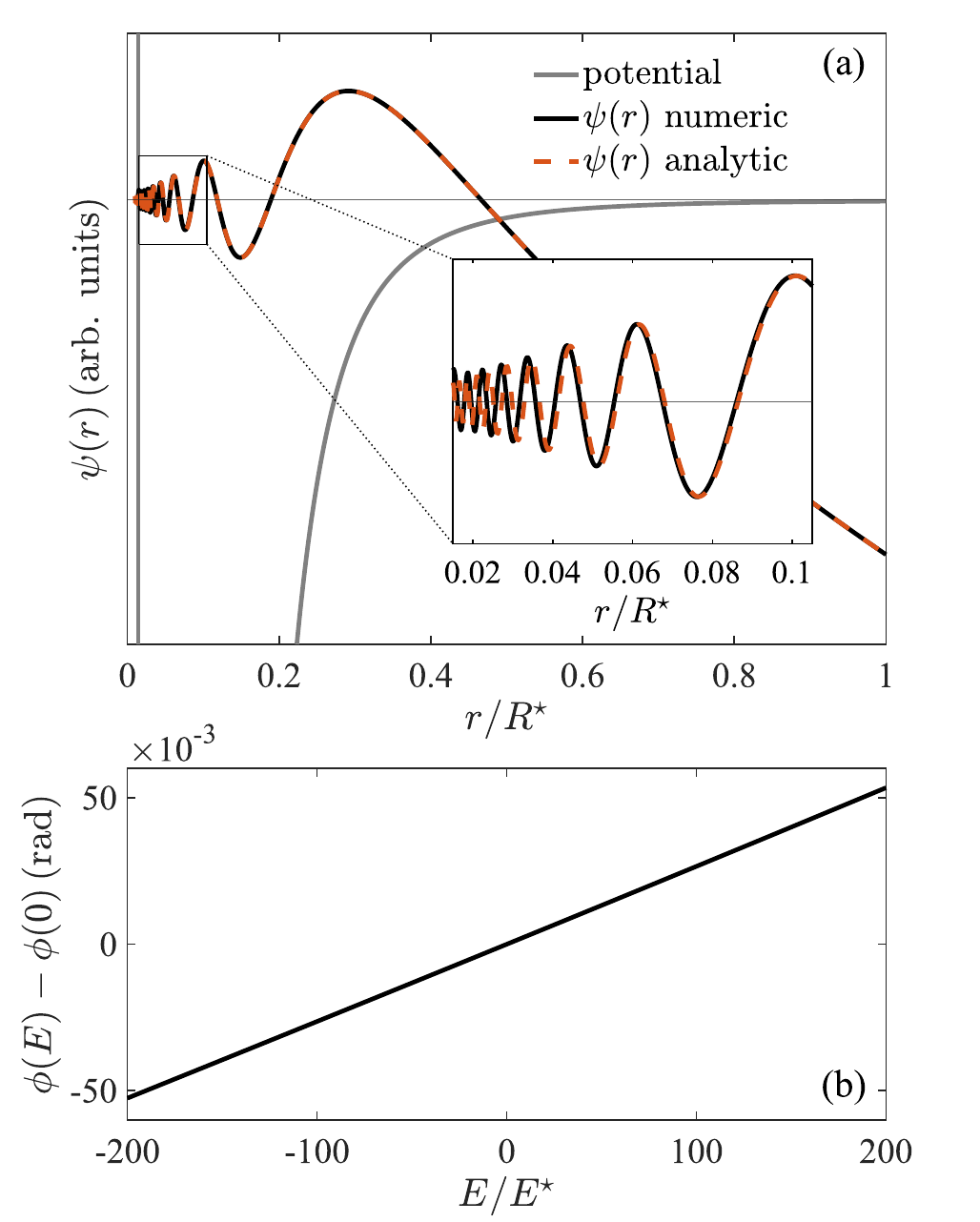}\caption{\textbf{(a)} A comparison
between the full numerical solution (solid black curve) to the 2-body
Schro\"{o}dinger equation and the analytic solution (dashed red curve) from
Eq.~\eqref{Eq:2B_ZEsol} with the amplitude and phase fit using the long-range
tail of the numerical solution. The atom-ion interaction potential is shown as
a gray curve for reference. A zoomed in region near the origin where the
analytic solution and the numerical begin to differ is shown in the inset. \textbf{(b)}
The phase $\phi$ is shown in radians as a function of scattering energy compared to the
zero energy phase.}%
\label{fig:phase_fit}%
\end{figure}

The strongly interacting limit is achieved at values of $r$ large enough that the $r^{-4}$ long-range tail of the atom-ion interaction dominates over the $r^{-6}$ repulsive core. In this limit the 2-body
Hamiltonian becomes purely that of the charge interacting with an induced
dipole%
\[
H_\mathrm{2b}\longrightarrow-\frac{\hbar^{2}}{2\mu_\mathrm{2b}}\frac{d^{2}}{dr^{2}}%
-\frac{C_{4}}{r^{4}},
\]
and the exact zero energy solution is known~\cite{idziaszek2011multichannel}:%
\begin{equation}
\psi\left(  r\right)  \propto r\sin\left(  \frac{R^{\ast}}{r}-\phi\right).
\label{Eq:2B_ZEsol}%
\end{equation}
The phase $\phi$ is determined by the short-range interactions, or alternatively fixed by the physical scattering length. In
the regime of scattering energies of interest here ($E\lesssim10\,\hbar\omega
$), the short range, finite energy solution to the two-body SE,
\begin{equation}
\hat{H}_\mathrm{2b}\psi=E\psi, \label{Eq:2body_SE}%
\end{equation}
can be matched to Eq.~(\ref{Eq:2B_ZEsol}) as long as $C_{6}r^{-6}\ll
C_{4}r^{-4}.$ Figure~\ref{fig:phase_fit}(a) shows the numerical solution to
Eq.~\ref{Eq:2body_SE} for the ${}^{171}\text{Yb}^{+}+{}^6\text{Li}$ system compared to Eq.~(\ref{Eq:2B_ZEsol}) with the phase set to $\phi=0.9983$, extracted by fitting
from the long-range tail of the numerical solution. As it is readily apparent, the two wavefunctions agree well when the $r^{-4}$ polarization potential
is dominant, and they become indistinguishable in the long-range regime. In Fig.~\ref{fig:phase_fit}(b) we show the 
energy-dependent phase compared to its zero-energy value. The phase was by fitting the numerical wavefunction to the Eq.~(\ref{Eq:2B_ZEsol}) over the $0.3 a_{\text{ho}}<r<0.8 a_{\text{ho}}$. The range of energies
here corresponds to approximately $450\,\hbar\omega$ for 
%a Paul trap with secular frequency $\omega = 2\pi \times 320 \,\text{kHz}$.
a typical Paul trap
secular frequency of $\omega\sim2\pi\times1$ MHz. 
Even over this large range
of energies, the phase varies on the $\sim 50$ milliradian level.  
We solve the Schr\"{o}dinger
equation in the classically allowed regime to emphasize the weak energy
dependence of the phase.

The limiting behavior of $H_\mathrm{ad}$ when $\theta \rightarrow \theta_\mathrm{c}$, and the weak energy dependence of the phase $\phi$, allow for cumulative effects of the deep short-range interaction potential in the adiabatic Schr\"{o}dinger equation in Eq.~(\ref{Eq:ad_SE}) to be absorbed into a single boundary
condition on $\Phi_n(R;\theta)$ at $\theta$ nearby $\theta_{c}$. This procedure is indeed inspired by quantum defect theory. Specifically, after extracting
$\phi$ from the free-space, zero-energy SE, we impose a node in the
angular wavefunction that coincides with a node in the zero-energy, 2-body solution, Eq.~(\ref{Eq:2B_ZEsol}). Each half oscillation included in the 2-body solution
corresponds to the inclusion of a molecular-ion bound state in the adiabatic
potentials starting from the most weakly bound state. Thus, we may systematically study the adiabatic potentials as more bound states are included.

We also note here that the Hamiltonian in Eq.~\eqref{Eq:H_ad} is invariant under
the parity transformation $\left\{  x,y\right\}  \longrightarrow\left\{
-x,-y\right\}  $ or equivalently $\theta\longrightarrow\theta+\pi$. Further,
due to the hard wall given by the $r^{-6}$ short-range repulsion in
\ $V_\mathrm{int}$, the even and odd parity states are degenerate. For each value
of the polar radius, $R$, we therefore limit ourselves only to solving the
adiabatic SE on the angular interval $\theta_{c}+\pi+\theta_{0}\leq\theta
\leq 2\pi + \theta_{c}-\theta_{0}$ where we enforce a node (corresponding to a hard wall boundary condition) at the angular position
\begin{equation}
\theta_{0}=\frac{R^{\ast}}{\sqrt{\beta
}R\left(  n_b\pi+\phi\right)}.\label{Eq:BC_pos}
\end{equation}
We also restrict the phase to the
range $0<\phi\leq\pi$. The position of the imposed angular boundary condition
also depends on the number of nodes included in the zero-energy two-body solution, given here by $n_b+1$. Hence, the number of weakly bound two-body molecular-ion states included is $n_b$. We note here that the small-angle
approximation used to extract Eq.~\eqref{Eq:H_ad_limit} means that the above approach breaks down at small radii, and a more complete description of the short-range interaction is needed. To avoid this, we will limit ourselves to potentials and scattering events that happen at ranges beyond this lower limit. To explore the small-$R$ regime, the two-body boundary condition must be implemented at shorter range with more deeply bound states incorporated.
However, including more than $n_b\sim4$ bound states becomes numerically infeasible.
Exploration of the high-energy, short-range interaction is a topic of ongoing investigation, and is beyond the scope of
this work.  

Here we extract $\phi$ from the numerical solution to the zero-energy 2-body Schr\"odinger equation with the interaction potential of Eq.~(\ref{Eq:2body_SE}).  Once $\phi$ and the boundary condition of Eq.~(\ref{Eq:BC_pos}) have been set, no further reference to the short-range interaction potential need be made, and in fact from this point forward, we omit if from the interaction and simply use $V_{\text{int}}=C_4/r^4$. The $C_6$ term in the interaction potential only serves to regularize the short-range behavior of the interaction and to provide a minimum length scale at which Eq.~(\ref{Eq:2B_ZEsol}) is still valid. Alternatively, the QDT phase could be fit to the atom-ion scattering length (or other known low energy scattering observables) of the free-space atom-ion system.

\subsection{Scattering Boundary Conditions}
\label{sec:scatbc}

As described in Sec.~\ref{sec:adiabaticrep}, we employ the adiabatic hyperspherical representation~\cite{macek1968pas}, in which the Schrodinger equation is reduced to a set of coupled equations in the hyperradius $R$ of the form Eq.~(\ref{Eq:rad_SE}).
At sufficiently large $R$, we may neglect off-diagonal elements of the nonadiabatic couplings $P_{mn}(R)$ and $Q_{mn}(R)$, such that the radial functions $F_n(R)$ satisfy an effective equation
\begin{equation}
\label{eq:asymse}
-\frac{\hbar^{2}}{2\mu}\frac{d^{2}%
}{dR^{2}}  F_{n}\left(  R\right)  +U_{\text{eff},n}\left(
R\right)  F_{n}\left(  R\right) = EF_{n}\left(  R\right),
\end{equation}
where the effective potential is
\begin{equation}
U_{\text{eff},n}(R) = U_n(R)-\frac{\hbar^2}{2\mu}\left(\frac{1}{4R^2}+Q_{nn}(R)\right)\label{Eq:ueff}.
\end{equation}
As shown in Appendix~\ref{appen:asymp}, the asymptotic forms of $U_n(R)$ and $Q_{nn}(R)$ together precisely cancel the $R^{-2}$ term in Eq.~(\ref{Eq:ueff}) that arises from the removal of first-derivatives in the radial kinetic energy. Solutions to Eq.~(\ref{eq:asymse}) at large $R$ serve to define the reactance $K$-matrix:
\begin{equation}
    F_{n}^{(\ell)}(R) \rightarrow f_{n}^{(\ell)}(R)\delta_{nm} - g_{n}^{(\ell)}(R) K_{nm}^{(\Pi)},\label{Eq:Kmat}
\end{equation}
where $\{f_n,\ g_n\}$ form a pair of energy-normalized, linearly independent solutions in energetically open channels. They are simple trig functions that we write as
\begin{align}
    % f_m&=\sqrt{\frac{\mu}{2\pi k_m R}}\sin{\left(k_m R+\frac{P_\ell \pi}{2}\right)}\\
    % g_m&=\sqrt{\frac{\mu}{2\pi k_m R}}\cos{\left(k_m R+\frac{P_\ell \pi}{2}\right)}
    f_{n}^{(\ell)}(R)&=\sqrt{\frac{\mu}{2\pi k_n}}\cos{\left(k_n R-\frac{\ell \pi}{2}\right)} \label{eq:regf}\\
    g_{n}^{(\ell)}(R)&=\sqrt{\frac{\mu}{2\pi k_n}}\sin{\left(k_n R-\frac{\ell \pi}{2}\right)}. \label{eq:irregg}
\end{align}
The reactance matrix $\mathbf{K}$ is related to the scattering matrix $\mathbf{S}$ in the usual way \cite{aymar1996multichannel},
\begin{equation}
\mathbf{S}^{(\Pi)}=\frac{1+i\mathbf{K}^{(\Pi)}}{1-i\mathbf{K}^{(\Pi)}},\label{Eq:SMat}
\end{equation}
and the scattering cross sections for $\Pi=\pm 1$ are computed as
\begin{equation}
    \sigma_{n'\leftarrow n}^{(\pm)}=\frac{1}{4}\frac{k_{n'}}{k_n}\left\vert S_{n'n}^{(\pm)}\mp (-1)^{n}\delta_{n'n} \right\vert^2. \label{Eq:crsossec}
\end{equation}

In order to understand the connection to the one-dimensional continuum, and the introduction of the parity-dependent parameter $\ell$, we must consider the noninteracting reference wavefunction. In the absence of an atom-ion interaction, eigenstates of Eq.~(\ref{Eq:H_ai}) may be written in the form
\begin{equation}
    \Psi_{\text{NI}}(x,y)=\mathcal{F}_\ell(y)\Upsilon_n(x), \label{Eq:psiNI}
\end{equation}
where $\Upsilon_n(x)$ are eigenstates of $H_\mathrm{i} = -\frac{\hbar^2}{2\mu} \pdd{}{x}+\frac{1}{2}\mu \omega^2 x^2$, and $\mathcal{F}(y)$ are eigenstates of $H_\mathrm{a} = -\frac{\hbar^2}{2\mu}\pdd{}{y}$. Since the noninteracting Hamiltonian $H_{\text{NI}}=H_i+H_a$ commutes with the parity operator $\hat{\Pi}=\hat{\Pi}_x\hat{\Pi}_y$, as well as with the individual reflection operators $\hat{\Pi}_x$ and $\hat{\Pi}_y$, the functions in Eq.~(\ref{Eq:psiNI}) satisfy
\begin{align}
    \hat{\Pi}_x\Upsilon_n(x)&=(-1)^n\Upsilon_n(x) \label{Eq:pixsym} \\
    \hat{\Pi}_y\mathcal{F}_\ell(y)&=(-1)^\ell \mathcal{F}_\ell(y) \label{Eq:piysym}\\
    \hat{\Pi}\Psi_{\text{NI}}&=(-1)^{n+\ell}\Psi_{\text{NI}}. \label{Eq:pisym}
\end{align}
The functions $\mathcal{F}_\ell(y)\propto\cos{(k y-\ell \pi/2)}$, where $\ell=0$ and $\ell=1$ represent the even and odd one-dimensional “partial waves”, respectively. Whether the “regular” solution $f_{n}^{(\ell)}$ for a given channel $n$ behaves as a cosine or a sine is determined by the total parity $\Pi$ according to $\ell = (1-\Pi (-1)^n)/2$, which ensures that Eq.~(\ref{Eq:pisym}) is satisfied in the noninteracting limit.

Let us now consider the fully interacting solutions to Eq.~(\ref{Eq:H_polar}) in the asymptotic limit. As $R\rightarrow \infty$, all relevant low-energy eigenstates of Eq.~(\ref{Eq:H_ad}) are localized to a narrow region near the $y$-axis, in the vicinity of $\theta \approx \frac{\pi}{2}, \frac{3\pi}{2}$.  Defining an angle $\varphi$ such that $\varphi = \pi/2-\theta$ for $y>0$ and $\varphi = \theta - 3\pi /2$ for $y<0$, we may now approximate $x\approx R\varphi$ and $|y| \approx R$ for all solutions belonging to relevant collision channels as $R\rightarrow \infty$. In this limit, the channel functions asymptotically approach the oscillator functions:
\begin{equation}
    \label{eq:asymphi}
    \Phi_n(R,\varphi)\rightarrow \sqrt{R}\;\Upsilon_n(R\varphi), 
\end{equation}
but with a two-fold degeneracy associated with the total parity $\Pi = \pm 1$. The degeneracy arises due to the hard-wall boundary conditions imposed near the coalescence angle. However, because our calculations are restricted to one side of the hard wall, we find only half of the solution set and have, therefore, a one-to-one correspondence in Eq.~(\ref{eq:asymphi}).

\section{Results}
\label{sec:results}
\subsection{Adiabatic Potentials}
\begin{figure}
    \hypertarget{fig:potentials}{}
    \centering
    \includegraphics[width=\linewidth]
    {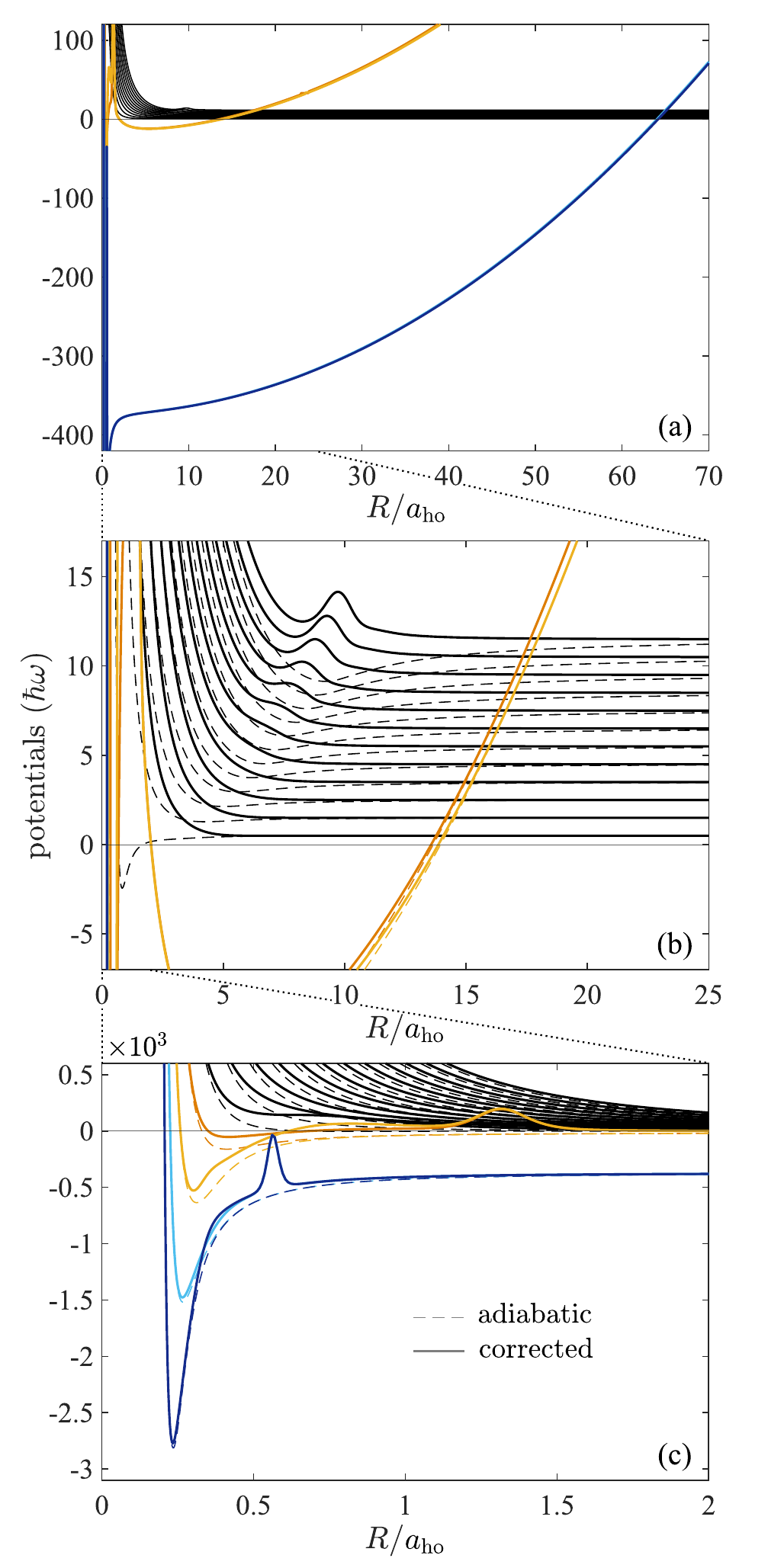}
    \caption{The adiabatic potentials are shown over in different ranges as a function of R. \textbf{(a)} The molecular-ion channels (yellow and blue) intersect the scattering channels (black) around $R\simeq15\,a_\mathrm{ho}$ and $R\simeq65\,a_\mathrm{ho}$. \textbf{(b,c)} The effective potentials $U_{\mathrm{eff},n}(R)$ (solid) are more repulsive than the bare adiabatic potentials $U_n(R)$ (dashed) due to the positive diagonal correction.}
    \label{fig:potentials}
\end{figure}

Here we calculate and analyze the adiabatic potentials for the ${}^{171}\text{Yb}^{+}+{}^{6}\text{Li}$ system. We set the QDT phase from Eq.~(\ref{Eq:2B_ZEsol}) to $\phi=0.9983$ in agreement with the zero-energy phase extracted from Eq.~(\ref{Eq:2body_SE}) using the interaction parameters from Ref.~\cite{trimby2022buffer}. We then set the angular boundary condition $\theta_0$ by including three molecular-ion bound states, i.e.~setting $n_b=3$ in Eq.~(\ref{Eq:BC_pos}). As more bound states are included, the depth of the more weakly bound states increases. Including more bound states changes the results discussed below slightly quantitatively, but the qualitative behavior is unaffected. Unless stated otherwise all potentials are found for three bound states, but only the weakest two are included for any scattering calculations. This procedure improves the convergence properties of scattering calculations sensitive to bound states attached to the shallowest two molecular-ion  

The results are shown in Fig.~\ref{fig:potentials}. The bare adiabatic potentials, $U_n(R)$, are shown as dashed curves while the effective potentials, $U_{\mathrm{eff},n}(R)$, are shown as solid curves.  Note that, as is required, the diagonal correction is positive, making the effective potentials more repulsive.
The black curves in Fig.~\ref{fig:potentials} are asymptotically flat.  Close inspection shows that each of these approaches an oscillator energy, $U_n(R)\rightarrow\hbar \omega(n+1/2)$.  These can be interpreted as the scattering potentials for a free Li atom interacting with a Yb$^+$ ion in the $n$th trap state. In the case of the lowest entrance scattering channel (corresponding asymptotically to the $n=0$ ion oscillator state), what appears at first as an attractive well becomes entirely repulsive with the diagonal correction. This small-$R$ repulsion is a result of the extra kinetic energy that comes from confining the system to smaller sizes. According to the Heisenberg uncertainty principle, as the size of the system is squeezed, the uncertainty in the momentum (and, correspondingly, the systems kinetic energy) must increase.  

In Fig.~\ref{fig:potentials}(\hyperlink{fig:potentials}{a}) and (\hyperlink{fig:potentials}{b}) we can see that a set of molecular-ion potentials, shown as yellow and blue curves, emerge cutting through the atom-ion scattering potentials in a series of very narrow avoided crossings. Asymptotically, these diabatic curves are quadratic and approach $U(R)\rightarrow \mu \omega^2 \cos^2(\theta_\mathrm{c})R^2/2-E_\mathrm{b}$, where $E_\mathrm{b}$ is the binding energy of the molecular-ion state. This leads us to interpret these as the potentials associated with the molecular-ion states that are experiencing the trapping potential. %Each molecular-ion bound state has two trapping potentials associated with it.  These two potentials can roughly be thought of as corresponding to the even and odd parity trap states.  
While the molecular-ion states are quadratic at large $R$ [see Fig.~\ref{fig:phase_fit}(\hyperlink{fig:potentials}{a})] as expected,  a nontrivial structure arises at small $R$, as shown in Fig.~\ref{fig:potentials}(\hyperlink{fig:potentials}{c}).  For both molecular-ion states, there are potential wells at small $R$. This indicates that, as the system is artificially compressed to small RMS sizes, the extra localizing potential associated with the ion trap can possibly create a molecular-ion state with binding energy below that of free space. For each of the molecular-ion potentials, the diagonal correction creates an effective barrier at $R\sim\left<r\right>_\mathrm{d}$ where $\left<r\right>_\mathrm{d}$ is the average size of the corresponding free-space molecular-ion vibrational state.  These diabatic states support a series of trapped molecular-ion states.  When the incident scattering energy in one of the scattering channels is close to one of these trap states, the atom can temporarily bind to the ion.  The binding energy that is released leaves the molecular-ion in a high-lying oscillator state associated with its center-of-mass motion. This dynamical process appears as a resonance in atom-ion scattering, induced by the confinement of the ion.

In truth, the crossings between the molecular-ion potentials and the  collision channels are very narrow avoided crossings. We have found that for any reasonable discrete grid of $R$ values, these crossing are narrower than the grid spacing and they may be treated as level crossings. The curves shown in Fig.~\ref{fig:potentials} are diabatized by simply reordering the potential indices to stay consistent through each crossing. In practice, this is done by choosing an index order that maximizes the overlap between channel functions at neighboring radial grid points. Nonadiabatic couplings are computed via Eqs.~(\ref{eq:pmat}) and (\ref{Eq:QviaP}) \emph{after} the diabatization process to ensure that the couplings are smooth functions of $R$. The asymptotic form of these potentials and that for the atom-ion collision channel potentials is derived in Appendix~\ref{appen:asymp}. At lowest order, the diabatic potentials are simply the molecular-ion binding energy plus the trapping potential, $U(R)\approx E_\mathrm{b}+\frac{1}{2}\mu \cos\theta_\mathrm{c} \omega^2 R^2$ where the extra factor of $\cos \theta_\mathrm{c}$ accounts for the slight mass difference between the $^{171}\text{Yb}^+$ ion and the $({}^{171}\text{Yb}{}^6\text{Li})^+$ molecular ion. For each bound state, there are two potentials.  Roughly, these two potentials can be thought of as corresponding to two trap states of opposite parity in the molecular-ion center of mass coordinate. Note that this is different than the total parity symmetry of the 2D system (under inversion of the ion and atom coordinates), which is conserved here, with degenerate even and odd total parity states.

\subsection{Nonadiabatic couplings}
Using the angular channel functions extracted at each radius, we can calculate the $P$-matrix through the Hellman-Feynman theorem as described in Sec.~\ref{sec:adiabaticrep}. The $Q$-matrix can then be calculated using Eq.~(\ref{Eq:QviaP}). Dynamically, transitions between different adiabatic channels are usually driven by the $P$-matrix via the derivative couplings in Eq~(\ref{Eq:rad_SE}). To get a qualitative understanding of the transition probability between two channels, $i$ and $j$, it is useful to consider the unitless coupling strength $$\mathcal{P}_{ij}=\frac{\hbar^2}{2\mu}\frac{\left|P_{ij}\right|^2}{\left|U_i(R)-U_j(R)\right|}.$$ The highest probability for a transition occurs at a peak in the coupling strength, which is often located at the closest point between two potentials going through an avoided crossing with each other. 

\begin{figure}
    \centering
    \includegraphics[width=\linewidth]{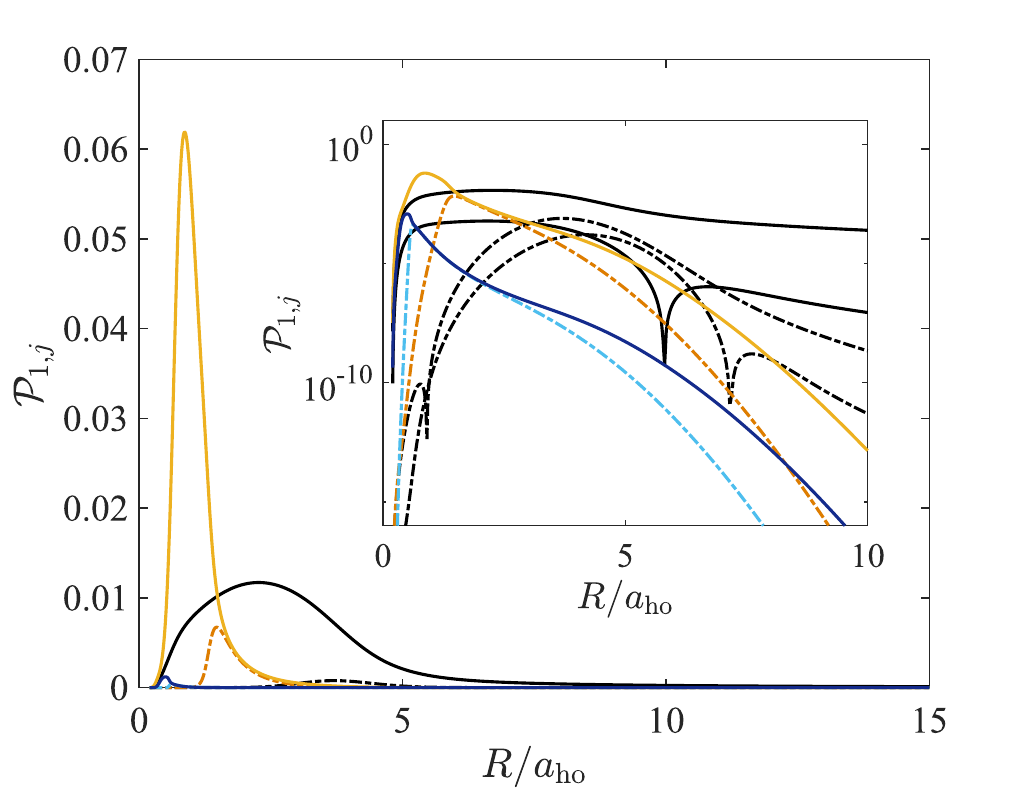}
    \caption{The coupling strength to the lowest scattering channel is shown as a function of $R$. Couplings to the lower and upper molecular-ion potentials are shown in solid yellow and dot-dashed orange respectively.  Couplings to the lower and upper second molecular-ion state potentials are shown as solid dark and dot-dashed light blue curves respectively.  Couplings to scattering channels with even parity (odd) ion trap states are shown as solid (dot-dashed) black curves. Inset: The same plot is shown on a log scale to make the weaker couplings visible.}
    \label{fig:couplings}
\end{figure}

Figure~\ref{fig:couplings} shows the coupling strength between the lowest scattering channel and the other adiabatic potentials plotted as a function of $R$.  The inset shows the same plot but on a log scale. It is immediately apparent that the lower of the two potentials associated with the most weakly bound state has the largest coupling strength to the first scattering channel. The coupling between the two molecular bound state potentials (not shown) is of the same order of magnitude as the lower potential to the collision channels. As a consequence, we expect the resonances associated with these trapping potentials to be broad.  On the other hand, the coupling strength to the second molecular-ion state potentials are two orders of magnitude smaller, indicating that any resonances associated with those potentials will likely be much narrower. We can also note that the peak in the coupling strength to the molecular-ion potentials occurs when $R$ is approximately the RMS size if the molecular-ion state, i.e. at $R\sim\left<r\right>_\mathrm{d}$ where $\left<\cdot\right>_\mathrm{d}$ indicates an expectation value of a molecular vibrational state in relative radial coordinate.  In the case of the second molecular-ion state, these peaks happen well within the repulsive barrier created by the diagonal correction in the scattering channel potentials and can only be accessed by tunneling.  This means that coupling to these state will be suppressed at low energies, and the associated resonances will broaden with increasing energy.

Among the scattering channels, due to the symmetry properties of the channel functions $\Phi_n(R;\theta)$, the coupling between the lowest and second excited scattering channel is the largest, peaking at approximately $R\sim a_\mathrm{ho}$ with a long-range tail that falls of as $\mathcal{P}\sim 1/R^2$ (see Appendix~\ref{appen:P_collision}). Provided that the collision energy is large enough that the second excited channel is energetically available, the strongest inelastic loss from the lowest entrance channel will be to the second excited channel. This pattern holds throughout the scattering channels with strong coupling to channels that correspond to ion trap states with $\pm 2$ oscillator quanta.

\subsection{Scattering}
With the adiabatic potentials and non-adiabatic $P$-and $Q$-matrices in hand, we can now analyze dynamics of this system. To calculate the scattering cross sections and transition probabilities, we employ the eigenchannel $R$-matrix method in conjunction with a finite-element $R$-matrix propagator~\cite{greene1983atomic,burke1999theoretical} to find the $K$-matrix of Eq.~(\ref{Eq:Kmat}), which is then related to the scattering matrix and cross section through Eqs.~(\ref{Eq:SMat}) and (\ref{Eq:crsossec}).
%This can then be related to the standard scattering matrix ${\bf S}=\left(1+i{\bf K}\right)\left(1-i{\bf K}\right)^{-1}$.  
It is important to note that because there is a hard core barrier, the adiabatic potentials for total even and odd parity (the parity of the atom and ion) are degenerate.  If we assume that scattering calculations are performed on the even (odd) parity potentials, then the long-range scattering wave function as discussed in Section~\ref{sec:scatbc} must follow the same (opposite) parity of the ion trap state.
 
Within this model, the only allowed scattering processes are elastic ones in which the atom scatters from the trapped ion without changing its trap state, or inelastic in which the atom either deposits excess kinetic energy, exciting the ion in the trap, or the atom absorbs energy from the ion as it falls to lower trap states. While these are the only allowed outcomes in the long time limit, the scattering delay can vary over a large range. 
Specifically, if the atom experiences a narrow resonance, the scattering process can be quite slow, and the system rattles in a highly excited molecular-ion trap state until it finds its way back out, kicking the atom back into the continuum.

To distinguish the different scattering processes, we number the scattering channels starting with the lowest channel (that corresponding asymptotically to a free atom and an ion trapped in the ground oscillator state) labeled as 1. We then extract transition probabilities from the $S$-matrix as $P_{i\leftarrow j}=\left|S_{ij}-\delta_{ij}\right|^2$. For instance, $P_{1\leftarrow 4}$ is the transition probability for the system starting in the 4th scattering channel with the ion in the 3rd excited oscillator state and ending in the first channel with the ion in the ground oscillator state. In this process, to conserve energy, the atom must end with the three extra oscillator quanta of kinetic energy.

\begin{figure}
    \hypertarget{fig:cross_section}{}
    \centering
    \includegraphics[width=\linewidth]{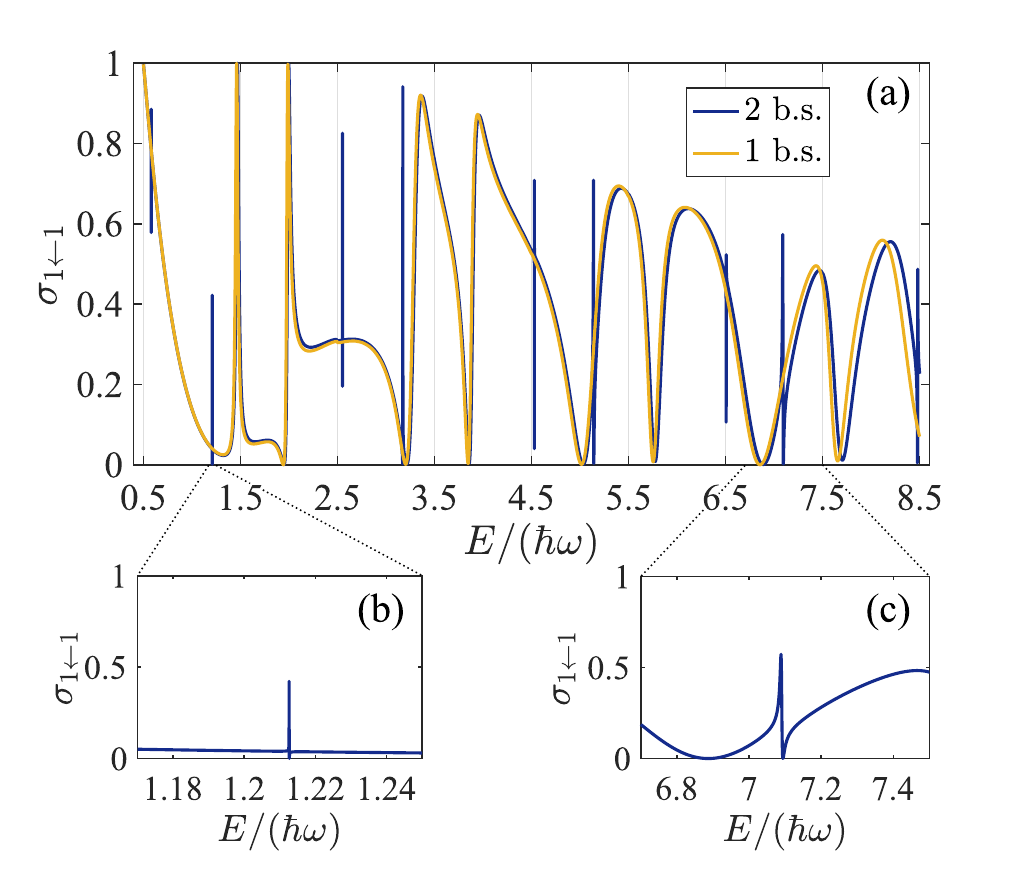}
    \caption{\textbf{(a)} The elastic cross section $\sigma_{1\leftarrow1}$ for two (blue) and one (yellow) bound states is shown as a function of collision energy. \textbf{(b,c)} Zoom on two resonances for the case of two bound states, highlighting the broadening at larger energies.}
    \label{fig:cross_section}
\end{figure}

% \subsubsection{Elastic Scattering}
\textit{Elastic Scattering –} Here we examine elastic scattering processes in which the kinetic energy of the atom is the same after scattering as it was before. Specifically, we look at elastic scattering of a Li atom scattering off of a Yb\textsuperscript{+} ion in the ground trap state (though the qualitative features would be the same if the ion were in an excited state).  The even total parity elastic scattering cross section in the lowest channel $\sigma^{(+)}_{1\leftarrow 1}$ of Eq.~(\ref{Eq:crsossec}) is shown in Fig.~\ref{fig:cross_section}(\hyperlink{fig:cross_section}{a}) as a function of total energy. Note that here we label the collision channels starting with $i=1$ which corresponds to the $n=0$ ion trap state. The atom's asymptotic kinetic energy is the total energy minus the trap energy of the ion, $KE=E-\hbar \omega/2$ in this case. The yellow curve here shows the cross section calculated including only the molecular-ion potentials associated with the most weakly bound molecular state. A clear set of resonances can be seen, each one corresponding to a molecular-ion trap state. The resonances arrive in pairs, one pair for each of the molecular-ion potentials, and are approximately evenly spaced with a new pair appearing for every $2\hbar \omega$ change in energy. Increasing the atom's kinetic energy allows it to reach smaller $R$ values before being repelled in the lowest effective scattering potential, increasing the probability of reaching peak coupling strength to molecular-ion potentials. This leads to broadening in the resonances with increasing energy, as expected.

The blue curve in Fig.~\ref{fig:cross_section}(\hyperlink{fig:cross_section}{a}) shows the same cross section calculation including two molecular-ion bound states. At low energy, the broad resonances associated with the first molecular-ion state are unchanged; however, a series of very narrow resonances corresponding to highly excited trap states of the second molecular-ion bound state appear. Like the broader resonances resulting from the shallower molecular-ion state, these narrow resonances appear in pairs, with new pairs arriving every $2\hbar \omega$ in energy.  Figures~\ref{fig:cross_section}(\hyperlink{fig:cross_section}{b}) and (\hyperlink{fig:cross_section}{c}) show two of the narrow resonances over a much tighter energy scale, highlighting that they too increase in width as scattering energy increases. We can see that at higher energies, the more deeply bound molecular potentials are coupling to the system enough to create noticeable shifts in the broad resonance positions.

We surmise that as deeper molecular-ion bound states are included, this pattern of resonance pairs will continue with narrower and narrower resonances being introduced in the low-energy cross section. Because they are trap states, at least in the low energy scattering, we expect the resonances pairs to be evenly spaced by $2\hbar \omega$ in energy. This behavior is unexpected in a system that is classically chaotic. The Bohigas-Giannoni-Schmit (BGS) conjecture, which states that the quantum mechanical spectrum of a classically chaotic system should follow a Wigner-Dyson distribution (WD) in its energy level spacing \cite{bohigas1984characterization}, would lead us to expect the resonances induced by the presence of the trapped molecular-ion state should follow the same WD distribution. Here, the spacing between resonant states is regular with only a few molecular-ion states contributing to the system while resonances associated with deeper states become nearly vanishingly narrow. However, at higher scattering energies, the effective coupling between potentials will become larger; the coupling between molecular-ion potentials at small $R$ will shift the resonance positions, and the regular spacing will likely disappear. The emergence of quantum chaotic scattering at higher energies in this system is a question of active interest, but is beyond the scope of this study.

% \subsubsection{Inelastic Scattering}
\begin{figure}
    \centering
    \includegraphics[width=\linewidth]{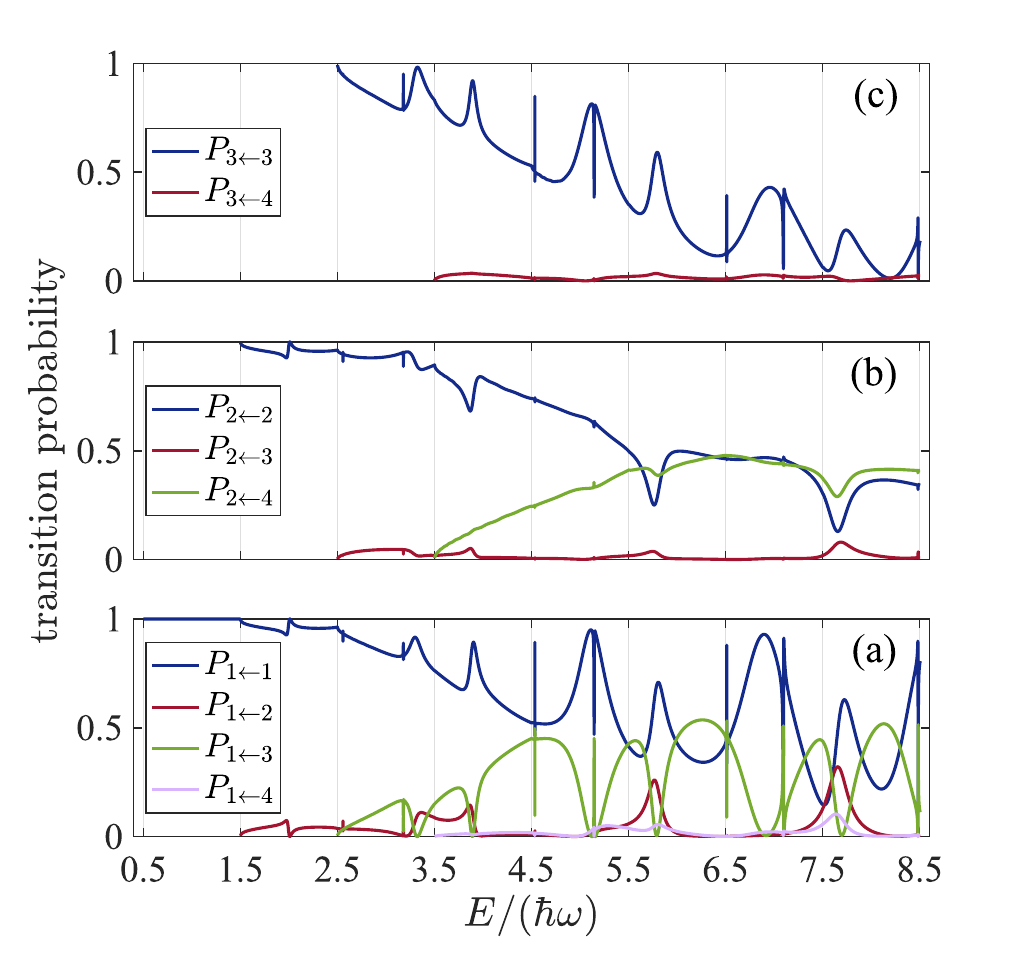}
    \caption{Transition probabilities to the \textbf{(a)} first, \textbf{(b)} second,  and \textbf{(c)} third channel are shown as a function of collision energy.}
    \label{fig:transitions}
\end{figure}
\textit{Inelastic Scattering –} When the scattering energy becomes greater than the threshold energy of the second scattering state, transitions to or from that state are possible. As each new threshold opens, new inelastic scattering pathways become available. Physically, these processes correspond to either (1) the atom depositing kinetic energy into the ion, leaving the ion in a higher oscillator state, or (2) the atom absorbing energy, leaving the ion in a lower oscillator state.  The relative rates at which the atom deposits or receives energy from the ion are of great interest for experimental efforts in sympathetic cooling. Figure~\ref{fig:transitions} shows the probabilities of exiting in channel $i$ given the system starts in channel $j$ as a function of energy. Here, we focus on scattering events in which the atom either keeps or increases its kinetic energy with transitions into the first (bottom panel (a)), second (middle panel (b)) and third (top panel (c)) channels. Notice that as the scattering energy increases, new channels become open and more pathways become available for scattering into the given state. When the scattering energy is near that of a molecular-ion trap state (the resonances seen in Fig.~\ref{fig:cross_section}), a corresponding resonance emerges in the transition probabilities. 

\section{Summary and Outlook}
\label{sec:summary}
We have created fully quantum mechanical description of scattering of a single low energy ${}^6$Li atom from a harmonically trapped $^{171}\text{Yb}^+$ ion in one dimension.  To overcome the vast difference in energy scale between the interaction energy, the secular trapping energy, and scattering energies, we employed a method inspired by the frame transformation approach of QDT. Specifically, short-range properties of the interaction potential were reduced to a single phase parameter extracted from the zero-energy wavefunction of the atom-ion system in free space. For the purposes of this work, the phase was fixed using the model interaction potential of \cite{trimby2022buffer}, and the resulting angular boundary condition was imposed to admit three molecular-ion bound states.  Exploring the behavior of the system across all possible quantum defect phases, $0\le \phi\le \pi$, is a topic of current interest. 

By harnessing the power of the adiabatic hyerpspherical method, a series of adiabatic potentials emerge. Asymptotically, these potentials can be broken into two categories: (1) potentials representing a free atom scattering from a trapped ion and (2) potentials associated with harmonically trapped molecular-ion states. We have studied the collisional resonances that arise from couplings between these two classes of potentials.

Classically, at low energies this system exhibits chaotic scattering behavior in which long-lived atom-ion complexes form.  In the quantum system, these complexes emerge as resonances in the scattering cross section caused by highly excited states of trapped molecular ions. At low energies, each molecular bound state contributes a series of pairs of resonances that are evenly spaced by $2\hbar \omega$. Resonances associated with deeper bound states are considerably narrower due to a combination of weaker non-adiabatic coupling to the scattering channels and tunneling suppression due to short-range repulsion in the effective potentials. The regularity of the low-energy spectrum belies what would be expected from the BGS conjecture on quantum chaos.  At higher energies, the resonances broaden, and coupling between molecular-ion states appears to start perturbing the spacing in the resonance positions. Scattering at higher energy and the possible emergence of the WD distribution in the spacing between resonances is an ongoing question of future inquiry.

\begin{acknowledgments}
N.P.M.~and S.T.R.~acknowledge support in part from the National Science Foundation through Grant NSF PHY-2409110 and PHY-2409111. N.P.M. and S.T.R. also acknowledge NSF support through ITAMP, from Quantum Delta NL through the University of Amsterdam, and in part by grant NSF PHY-2309135 to the Kavli Institute for Theoretical Physics (KITP).  A.S.N. was supported by the Dutch Research Council (NWO/OCW) as a part of the Quantum Software Consortium (project number 024.003.037), Quantum Delta NL (project number NGF.1582.22.030) and ENW-XL grant (project number OCENW.XL21.XL21.122).
\end{acknowledgments}

\appendix
\section{Asymptotic forms of potentials and couplings}\label{appen:asymp}
In this appendix, we derive the asymptotic forms for the scattering potentials, the molecular-ion potentials, and the various associated $P$-matrix elements. 
We note that the channel functions corresponding to the scattering channels and those corresponding to the molecular-ion potentials are strongly localized in the angular coordinate in the large-$R$ limit.  This means that the P-matrix elements between these two categories of states will be exponentially suppressed as $R\rightarrow\infty$. 

We shall use oscillator units for convenience, scaling all masses by $m_\mathrm{i}$, energies by $\hbar\omega$, and lengths by $a_\mathrm{ho}=\sqrt{\hbar/(m_\mathrm{i} \omega)}$. Note that in these units, the reduced mass becomes $\mu=\sqrt{m_\mathrm{a}/m_\mathrm{i}}=\tan{\theta_\mathrm{c}}$ and $\beta=(1+\mu^2)/\mu$.

\subsection{Collision-channel potentials}
In the $R\rightarrow \infty $ limit, our collision channels correspond to a free atom far from the trapped ion, and the adiabatic potentials approach the oscillator energies of the trapped ion.  The corresponding channel functions $\Phi_n(R;\theta)$ can be expressed in terms of oscillator solutions localized in the vicinity of $\theta=\pm\pi/2$. For convenience, let us concentrate on the region near $\theta\approx\pi/2$ by defining $\varphi = \pi/2 -\theta$.  Symmetries of the channel functions discussed in Sec.~\ref{sec:scatbc} can be used to consider the region $\theta \approx -\pi/2$. Neglecting the vanishing atom-ion interaction, but retaining the full form of the trapping potential for the ion, the adiabatic Hamiltonian in this limit becomes
\begin{equation}
\lim_{R\rightarrow \infty} H_\mathrm{ad} = -\frac{1}{2\mu R^2} \pdd{}{\varphi}+\frac{1}{2}\mu R^2\sin^2{\varphi}
\end{equation}
for $\varphi\ll 1$, we let $\rho = \sqrt{\mu} R \varphi$: 
\begin{equation}
\label{eq:hasymho}
    \lim_{R\rightarrow\infty}H_\mathrm{ad}\approx-\frac{1}{2} \pdd{}{\rho}+\frac{1}{2}\rho^2.
\end{equation}
The harmonic potential in $z_\mathrm{i}$ is anharmonic in $\varphi$, but we can use perturbation theory to capture this anharmonicity at large $R$ since it is small.  We define the perturbation as
\begin{equation}
    V_{\text{pert}}=\frac{1}{2}\mu R^2\left(\sin^2{\varphi}-\varphi^2\right).
\end{equation}
The correction to the unperturbed energy in the asymptotic limit of $U_n^{(0)} = (n+1/2)$ (with $n$ starting at $n=0$ here for convenience) is computed as the expectation value:
$U_n^{(1)}=\left\langle n \right|V_{\text{pert}}\left|n \right\rangle$, which immediately reduces to
\begin{equation}
       U_n^{(1)}  =\frac{1}{4}\mu R^2\left[1 - \left \langle n  \left |\cos{(2\varphi)}  \right | n \right \rangle\right] - \frac{1}{2}\left(n+\frac{1}{2}\right).
\end{equation}
Ladder operators provide the most expedient path to evaluating the expectation value $ \left \langle n  \left |\cos{(2\varphi)}  \right | n \right \rangle$.  To do so, we note that $\cos{(2\varphi)}=(e^{2i\varphi}+e^{-2i\varphi})/2$, and that $2i\varphi = 2i\rho/\sqrt{\mu}R$ where $\rho = (a+a^\dag)/\sqrt{2}$ in light of Eq.~(\ref{eq:hasymho}).  Using the Baker-Campbell-Hausdorff formula~\cite{GriffithsQM}, it is straightforward to show that
\begin{equation}
    \left\langle n \left|e^{\pm 2i\varphi} \right| n\right\rangle = e^{t^2/2}\sum_{l=0}^n{\frac{\left(t^{2}\right)^l}{l!}\binom{n}{l}}, 
\end{equation}
where $t = \pm \sqrt{2}i/\sqrt{\mu}R$, $t^2 = -2/\mu R^2$ and $\binom{a}{b}$ is the binomial coefficient $a$ choose $b$. The sum is precisely a Laguerre polynomial $L_n(-t^2)$, and so we have
\begin{equation}
    \left\langle n \left|e^{\pm 2i\varphi} \right| n\right\rangle = e^{\frac{-1}{\mu R^2}}L_n\left(\frac{2}{\mu R^2}\right).
\end{equation}
The expectation value of $\cos{2\varphi}$ then becomes straightforward, giving
\begin{align}
    U_n^{(1)}(R) =& \frac{1}{4}\mu R^2\left[1 -  e^{\frac{-1}{\mu R^2}}L_n\left(\frac{2}{\mu R^2}\right)\right] \notag\\
    & - \frac{1}{2}\left(n+\frac{1}{2}\right).
\end{align}
Isolating the leading order correction at large $R$ gives
\begin{equation}
    U_n^{(1)}(R)=-\frac{1+2n(n+1)}{8\mu R^2}.
\end{equation}
It's worth noting at this point that when the diagonal correction $-Q_{nn}(R)/(2\mu)$ (computed below) is included along with the first order correction from perturbation theory, the effective potential, $U_{\text{eff},n}(R)=U_n(R)-Q_{nn}(R)/(2\mu) -1/(8\mu R^2)$, acquires an asymptotic form behaving as $U_n(R)=n+1/2 + \mathcal{O}(R^{-4})$, appropriate for a two-body collision channel.

\subsection{\texorpdfstring{$P_{mn}(R)$}{Pnm(R)} and \texorpdfstring{$Q_{mn}(r)$}{Qnm(r)} for collision channels}
\label{appen:P_collision}
To compute the asymptotic forms of the nonadiabatic couplings, we first write the channel functions explicitly in the large-$R$ limit:
\begin{equation}
    \Phi_n(R;\theta)=\left(\frac{\mu R^2}{\pi}\right)^{1/4}\frac{1}{\sqrt{2^n n!}}H_n(\sqrt{\mu}R\varphi)e^{-\frac{\mu R^2 \varphi^2}{2}}
\end{equation}
After performing the radial derivatives, as prescribed in Eq.~(\ref{eq:pmat}), and subsequently using the relation $\rho\approx \sqrt{\mu}R\varphi$, the asymptotic form of the $P$-matrix element can be readily evaluated using ladder operators to give
\begin{align}
    P_{nm}(R) \rightarrow \frac{1}{2R} & \left[ \delta_{n,m-2} \;\sqrt{m(m-1)} \right.\notag \\
    & \left. - \delta_{n,m+2}\sqrt{(m+1)(m+2)}\right]
\end{align}
Similar methods can be used to evaluate asymptotic form of the second derivative coupling $Q_{nm}(R)$, although we find it more convenient to evaluate the symmetric matrix
\begin{align}
      \tilde{Q}_{nm}(R)&=\left\langle \pd{\Phi_n(R;\theta)}{R} \middle|\pd{\Phi_m(R;\theta)}{R}\right\rangle \nonumber\\
      &= - \left[P^2\right]_{nm}(R)
\end{align}
which is related to $Q_{nm}$ via Eq.~(\ref{Eq:QviaP}).  The result is
\begin{align}
    \tilde{Q}_{nm}(R) &= \delta_{nm}\left(\frac{2+m^2+m(3+|m-1|)}{4R^2}\right) \notag\\
    & - \delta_{n,m-4}\left(\frac{\sqrt{m(m-1)(m-2)(m-3)}}{4R^2}\right) \notag \\
    & - \delta_{n,m+4}\left(\frac{\sqrt{(m+1)(m+2)(m+3)(m+4)}}{4R^2}\right)
\end{align}
Couplings between collision channels and molecular-ion channels are exponentially suppressed at large $R$, so we neglect their asymptotic forms. The next section provides details on how we compute the asymptotic forms among the molecular-ion channels.
\subsection{Molecular-ion potentials}
Starting with the adiabatic Hamiltonian of Eq.~(\ref{Eq:H_ad}) in oscillator units of the ion, we rotate into a new angular coordinate $\alpha=\theta-\theta_\mathrm{c}$:
\begin{align}
\hat{H}_\mathrm{ad}=&-\frac{\hbar^2}{2\mu R^2}\frac{\partial^2}{\partial\alpha^2}+\frac{1}{2}\mu R^2\cos^2(\alpha+\theta_\mathrm{c}) \nonumber\\
&+V_\mathrm{int}(\sqrt{\beta}R\sin\alpha) \nonumber.
\end{align}  
In the limit where $R\gg R^*$ we can note that, at fixed $R$, the angular potential has two deep wells  near $\alpha=0$ and $\alpha=\pi$ corresponding to the atom-ion interaction potential. 
At large $R$, the molecular-ion state corresponds to angular solutions strongly localized near these two wells. Note that because the trap potential is now at an angle in the polar space, the two interaction wells are slightly offset from one another by the trap potential. The asymptotic angular channel functions that corresponds to the molecular-ion state is the bound states localized in $\alpha$ inside these two interaction regions:
 \begin{equation}
    \Phi^{(\pm)}_\mathrm{d}(R;\alpha)=A\psi_\mathrm{d}(r^{(\pm)}), \label{Eq:mol_st}
\end{equation}
where $r^{(+)}=\sqrt{\beta}R\sin\alpha$ and $r^{(-)}=\sqrt{\beta}R\sin(\pi-\alpha)$. Here $A$ is a radially dependent normalization constant and $\psi_\mathrm{d}(r)$ is a bound state solution to the two-body Schr\"odinger equation with binding energy $E_\mathrm{d}$:
\begin{equation}
    \hat{H}_\mathrm{2b}\psi_\mathrm{d}(r)=-E_\mathrm{d} \psi_\mathrm{d}(r).
\end{equation}

The normalization coefficient can be found by integrating in $\alpha$, i.e.
$$
1=\int_0^\pi \left| \Phi_\mathrm{d}^{(\pm)}(R;\alpha) \right|^2 d\alpha .
$$
At large $R$ the molecular-ion states will be highly localized so that we can use the small-angle approximation: $r^{(+)}\approx\sqrt{\beta}R\alpha$ and $r^{(-)}\approx\sqrt{\beta}R(\pi-\alpha)$.  Inserting this as a substitution gives
$$
1=\frac{1}{\sqrt{\beta}R}\int_0^\infty \left|A \right|^2 \left|\psi_\mathrm{d}(r) \right|^2 dr.
$$
Assuming that $\psi_\mathrm{d}$ is normalized in the relative coordinate, this gives $A=\sqrt{\sqrt{\beta}R}$.

In the $R\gg R^*$, we employ a Taylor series near the interaction wells for $r^{(\pm)}\ll R^*$ giving
\begin{equation}
    H_\mathrm{ad}\approx H_\mathrm{2b}+\frac{1}{2}\mu \cos^2\theta_\mathrm{c} R^2 \pm \frac{1}{2}\mu \frac{\sin 2\theta_\mathrm{c}}{\sqrt{\beta}} R r^{(\pm)}\label{Eq:Had_asymp},
\end{equation}
where $H_\mathrm{2b}$ is the free-space atom-ion Hamiltonian. Here, we have taken $r^{(+)}\approx \sqrt{\beta} \alpha$ and $r^{(-)}\approx \sqrt{\beta} (\pi-\alpha)$. Note that the quadratic potential in $R$ represents the trapping potential for the molecular ion. The extra factor of $\cos^2\theta_\mathrm{c}$ accounts for the extra mass in the molecular state, while the extra terms that are linear in $R$ account for the asymmetry of the trapping potential in the $\alpha$ angular coordinate for the $\Phi^{(\pm)}$ states. Therefore, the adiabatic potentials in the large-$R$ limit associated with the molecular-ion state are simply the expectation values of the adiabatic Hamiltonian in the $\Phi_\mathrm{d}^{(\pm)}$ states:
\begin{equation}
U_\mathrm{d}^{(\pm)}=\left<\Phi_\mathrm{d}^{(\pm)}\left|H_\mathrm{ad}\right|\Phi_\mathrm{d}^{(\pm)}\right>.
\end{equation}
The asymptotic channel function used here is an eigenfunction of the first two terms from Eq.~(\ref{Eq:Had_asymp}). The final expectation value can be performed using the substitution $r^{(\pm)}=\sqrt{\beta}R(\pi/2\mp\pi/2\pm\alpha)$ to give
$$
\left<\Phi_\mathrm{d}^{(\pm)}\left|\frac{1}{2}\mu \frac{\sin 2\theta_\mathrm{c}}{\sqrt{\beta}} R r^{(\pm)}\right|\Phi_\mathrm{d}^{(\pm)}\right>=\frac{1}{2}\mu \frac{\sin 2\theta_\mathrm{c}}{\sqrt{\beta}} R\left<r\right>_\mathrm{d},
$$
where $\left<r\right>_\mathrm{d}=\left<\psi_\mathrm{d}\left|r\right|\psi_\mathrm{d}\right>$ is the average radial size of the molecular ion in the corresponding vibrational state. Putting this all together, for each molecular-ion state, there are two adiabatic potentials that asymptotically go to
\begin{align}
U_\mathrm{d}^{(\pm)}(R)=-E_\mathrm{d}&+\frac{1}{2}\mu\cos^2\theta_\mathrm{c} R^2 \label{Eq:asymp_mol_pot}\nonumber\\ 
&\pm \frac{1}{2}\mu \frac{\sin 2\theta_\mathrm{c}}{\sqrt{\beta}}\left<r\right>_\mathrm{d} R.
\end{align}

\subsection{\texorpdfstring{$P_{d_1 d_2}(R)$}{Pd1d2(R)} for molecular-ion potentials}
\label{appen:P_molecular}
Here we derive the asymptotic form of the nonadiabatic P-matrix elements between different states. We can note that in the large-$R$ limit, the wavefunctions corresponding to the $+$ and $-$ molecular-ion potentials are highly localized and separated, meaning that any $P$-matrix element between them will be strongly suppressed. This means that asymptotically, only the derivative coupling matrix elements between the different molecular-ion channels need to be calculated. We proceed using the Hellmann-Feynman theorem of Eq.~(\ref{Eq:PElem}) written in terms of $\alpha=\theta-\theta_c$ in the $\alpha\ll 1$ limit:
\begin{align}
P_{d_1d_2}\left(  R\right)   &  =\frac{1}{R}\left[  4\frac{\left\langle \Phi
_{d_1}\left(  R;\theta\right)  \left\vert \frac{1}{2}\mu R^{2}\cos
^{2}(\theta_\mathrm{c})\right\vert \Phi_{d_2}\left(  R;\theta\right)  \right\rangle }%
{U_{d_2}\left(  R\right)  -U_{d_1}\left(  R\right)  }\right.  \label{Eq:PElem_mol}\nonumber\\
&  -\left.  2\frac{\left\langle \Phi_{d_1}\left(  R;\theta\right)  \left\vert
V_{int}\left(  \sqrt{\beta}R\alpha    \right)
\right\vert \Phi_{d_2}\left(  R;\theta\right)  \right\rangle }{U_{d_2}\left(
R\right)  -U_{d_1}\left(  R\right)  }\right],  
\end{align}
where $\Phi_{d_i}$ is given by Eq.~(\ref{Eq:mol_st}) for the $i$th molecular-ion state. The first term is zero by orthogonality. The remaining matrix element integral can be done using the substitution $r=\sqrt{\beta}R\alpha$. In the large-$R$ limit, this transforms the angular matrix element into an integral over the atom-ion separation coordinate, i.e.
\begin{align}
    \left<\Phi_{d_1}\left|V_{int}(\sqrt{\beta}R\alpha)\right|\Phi_{d_2}\right>=&\int_0^\infty \psi^*_{d_1}(r)V_{int}(r)\psi_{d_2}(r)dr \nonumber\\
    =& \left<\psi_{d_1}|V_{int}(r)|\psi_{d_2}\right>_\mathrm{d},
\end{align}
where $\langle\cdot\rangle_d$ indicates the matrix element is taken in the atom-ion separation coordinate. 

Evaluating the denominator in  Eq.~(\ref{Eq:PElem_mol}) and truncating to the quadratic term of Eq.~(\ref{Eq:asymp_mol_pot}) gives the final asymptotic form for the $P$-matrix element:
\begin{equation}
    P_{d_1d_2} =\frac{-2}{R}\frac{\left<\psi_{d_1}|V_{int}(r)|\psi_{d_2}\right>_\mathrm{d}}{E_{d_1}-E_{d_2}
    %\pm\frac{\mu R\sin 2\theta_\mathrm{c}}{2\sqrt{\beta}} \left(\left<r\right>_{d_2}-\left<r\right>_{d_1}  \right)
    },
\end{equation}
where $E_{d_i}$ is the binding energy of the $i$th molecular-ion state.
\bibliography{AtomIonRefs.bib}
\end{document}